%% file: main.tex
\documentclass[acmsmall,screen]{acmart}

\input{macro}

\begin{document}



\title{ClarifyGPT: Empowering LLM-based Code Generation with Intention Clarification}



 
 \author{Fangwen Mu}
 \authornote{Both authors contributed equally to this research}
 \affiliation{%
   \institution{Institute of Software, Chinese Academy of Sciences}
   \city{Beijing}
   \country{China}}
 \email{fangwen2020@iscas.ac.cn}

 \author{Lin Shi}
 \authornotemark[1]
 \affiliation{%
   \institution{Beihang University}
   \city{Beijing}
   \country{China}}
 \email{shilin@buaa.edu.cn}

 \author{Song Wang}
 \affiliation{%
   \institution{York University}
   \city{Toronto}
   \country{Canada}}
 \email{wangsong@yorku.ca}

 \author{Zhuohao Yu}
 \affiliation{%
   \institution{Institute of Software, Chinese Academy of Sciences}
   \city{Beijing}
   \country{China}}
 \email{yuzhuohao23@mails.ucas.edu.cn}

 \author{Binquan Zhang}
 \affiliation{%
   \institution{Beihang University}
   \city{Beijing}
   \country{China}}
 \email{binquan@buaa.edu.cn}

  \author{ChenXue Wang}
 \affiliation{%
   \institution{Institute of Software, Chinese Academy of Sciences}
   \city{Beijing}
   \country{China}}
 \email{chenxuew02@gmail.com}

  \author{Shichao Liu}
 \affiliation{%
   \institution{Software IDE innovation Lab, Huawei Central Software Institute}
   \city{Beijing}
   \country{China}}
 \email{liushichao2@huawei.com}

  \author{Qing Wang}
 \affiliation{%
   \institution{Institute of Software, Chinese Academy of Sciences}
   \city{Beijing}
   \country{China}}
 \email{wq@iscas.ac.cn}

\input{sec/abs}

\maketitle



\input{sec/intro-v2}
\input{sec/rw}
\input{sec/method}

\input{sec/exp}
\input{sec/result}
\input{sec/discussion}
\input{sec/conclusion}

\bibliographystyle{ACM-Reference-Format}
\bibliography{ref}
\end{document}

%% file: macro.tex
\usepackage{graphicx}
\usepackage{array}
\usepackage{amsmath}
\usepackage{algorithm}
\usepackage{hyperref}
\usepackage{multirow}
\usepackage{caption}
\usepackage{flushend}
\usepackage{balance}
\usepackage{ulem}
\usepackage{color, colortbl}
 \usepackage{enumitem}
 \usepackage{balance}
\usepackage{booktabs}
\usepackage{fancyhdr}
\usepackage[framemethod=TikZ]{mdframed}
\usepackage{tcolorbox}
\makeatletter
\newcommand{\mybox}[1]{%
  \setbox0=\hbox{#1}%
  \setlength{\@tempdima}{\dimexpr\wd0+13pt}%
  \begin{tcolorbox}[boxrule=0.5pt, colback=gray!10, arc=4pt,
      left=6pt,right=6pt,top=6pt,bottom=6pt,boxsep=0pt]
    #1
  \end{tcolorbox}
}

\usepackage{makecell}

\usepackage{float}
\usepackage{listings}

\definecolor{codegreen}{rgb}{0,0.6,0}
\definecolor{codegray}{rgb}{0.875,0.875,0.875}
\definecolor{codepurple}{rgb}{0.58,0,0.82}
\definecolor{backcolour}{rgb}{0.95,0.95,0.92}

\lstdefinestyle{mystyle}{
  aboveskip=2mm,
  belowskip=2mm,
  showstringspaces=false,
  columns=flexible,
  numbers=none,
  backgroundcolor=\color{backcolour},
  commentstyle=\color{codegreen},
 keywordstyle=\color{magenta},
    numberstyle=\tiny\color{codegray},
    stringstyle=\color{codepurple},
    basicstyle=\small\ttfamily,
    breakatwhitespace=false,         
    breaklines=true,                 
    captionpos=b,                    
    keepspaces=false,                 
    numbersep=5pt,                  
    showspaces=false,                
    showstringspaces=false,
    showtabs=false,                  
    tabsize=2,
    escapeinside=``
}
\newcommand{\tool}{\textsc{ClarifyGPT}}

%% file: sec/abs.tex
\begin{abstract}

{Large Language Models (LLMs), such as ChatGPT, have demonstrated impressive capabilities in automatically generating code from provided natural language requirements.
However, in real-world practice, it is inevitable that the requirements written by users might be ambiguous or insufficient. 
Current LLMs will directly generate programs according to those unclear requirements regardless of interactive clarification, which will likely deviate from the origin user intents.}
{To bridge that gap}, we introduce a novel framework named {\tool}, which aims to enhance code generation by empowering LLMs with the ability to identify ambiguous requirements and ask targeted clarifying questions. In particular, {\tool} first detects whether a given requirement is ambiguous by performing a code consistency check. If it is ambiguous, {\tool} prompts an LLM to generate targeted clarifying questions. After receiving question responses, {\tool} refines the ambiguous requirement and inputs it into the same LLM to generate a final code solution.
{To evaluate our {\tool}, we first conduct a human evaluation involving ten participants who use {\tool} for code generation on two publicly available benchmarks: MBPP-sanitized and MBPP-ET. The results show that {\tool} {elevates the performance (Pass@1) of GPT-4 from 70.96\% to 80.80\% on MBPP-sanitized.}
Furthermore, to perform large-scale automated evaluations of {\tool} across different LLMs and benchmarks without requiring user participation, we introduce a high-fidelity simulation method to simulate user responses. The automated evaluation results also demonstrate that {\tool} can significantly enhance code generation performance compared to the baselines. In particular, {\tool} improves the average performance of GPT-4 and ChatGPT across four benchmarks from 68.02\% to 75.75\% and from 58.55\% to 67.22\%, respectively.
We believe that {\tool} can effectively facilitate the practical application of LLMs in real-world development environments.}

\end{abstract}

%% file: sec/intro-v2.tex
\section{Introduction}
Code generation aims to produce a code snippet that satisfies the user's intent expressed in a natural language requirement. This task, which offers potential cost savings, accelerates programming activities, and facilitates software development, has consequently garnered attention across various domains, e.g., natural language processing, artificial intelligence, and software engineering. Recent efforts tackle this task by leveraging Large Language Models (LLMs) with billions of parameters, such as ChatGPT\cite{chatgpt} and CodeGen\cite{nijkamp2022codegen}. The LLMs take the natural language requirements (i.e., prompts) as inputs and output the corresponding code snippets, achieving remarkable progress in code generation. 

However, in real-world {practice}, {due to the diversity of user experience and perspective, it is inevitable that the requirements written by users might be ambiguous or insufficient.}
{For example}, 
the requirement \textit{"Write a function to sort a list of elements"} does not specify whether the user intends for the list to be sorted in ascending or descending order. 
Current LLMs do not handle such ambiguous requirements: they rarely ask users to clarify these requirements and instead directly generate programs that may deviate from the users’ needs~\cite{kuhn2023clam}.
{Current LLMs-based code generation approaches lack the mechanism of clarifying unclear requirements \cite{kuhn2023clam, krasheninnikov2022assistance}, i.e., they directly generate programs according to those unclear requirements regardless of interactive clarification.}
In contrast, when human developers encounter ambiguous requirements, they typically seek additional information by interactively asking clarifying questions to the users. For the above example, a simple clarifying question such as \textit{"Should the sorting be in ascending or descending order?"} could help disambiguate the requirement. 

In light of this observation, we argue that empowering LLMs with the ability to automatically ask clarifying questions for ambiguous requirements is necessary for improving the quality and efficiency of code generation. However, it is quite challenging to empower LLMs with this ability due to the following barriers.
\textbf{(1) When to Ask Clarifying Questions?} In practical development environments, numerous requirements exist, including both ambiguous and unambiguous ones. Failure to concentrate on questioning only the ambiguous requirements can lead to unnecessary interactions between LLMs and users regarding well-defined requirements. These unnecessary interactions, in turn, can diminish efficiency and compromise the user experience.
\textbf{(2) What Clarifying Questions Should be Asked?} The quality of clarifying questions also influences the efficiency and performance of code generation. Precisely and targeted questions aid users in expressing their intents clearly, ensuring that the obtained responses are directly relevant to the ambiguities present in the requirements. Vague or broad questions increase the risk of obtaining off-topic or irrelevant responses, potentially hindering LLMs from comprehending user intents.

In this paper, we propose a novel framework called {\tool} that enhances LLM-based code generation via requirement clarification.
{\textbf{First}, we employ a two-step code consistency check to decide when to ask clarifying questions. We are motivated by the observation that feeding a clear requirement to LLMs usually results in generating diverse code snippets that behave consistently, i.e., given the same test inputs, those different code snippets will likely return the same outputs. While feeding an unclear requirement, LLMs are likely to generate diverse code snippets that behave differently.
Specifically, in the first step, {\tool} aims to generate numerous high-quality test inputs for a given requirement via type-aware mutation. 
In the second step, {\tool} inputs the given requirement into an LLM to sample $n$ code solutions and checks whether they produce identical outputs when tested with the generated input. 
If the outputs are not identical, {\tool} determines that the requirement requires further clarification; and vice versa. 
} 
\textbf{Second}, we employ the reasoning-based prompting for generating clarification questions. Initially, {\tool} directs LLMs to analyze the factors contributing to the ambiguity of the given requirement by comparing code solutions with different functionalities. 
Subsequently, it formulates targeted clarifying questions based on the results of this analysis.
By comparing these different code implementations, potential points of ambiguity in the requirements can be readily identified.  After detecting the points of ambiguity in the requirements, the LLMs can generate targeted clarifying questions for them.
\textbf{Finally}, {\tool} refines the original requirement based on the generated questions and their responses and generates the final code solution.

To assess the effectiveness of {\tool}, we first integrate GPT-4~\cite{DBLP:journals/corr/abs-2303-08774} into {\tool} and recruit ten participants to evaluate its performance on two public benchmarks (MBPP-sanitized~\cite{DBLP:journals/corr/abs-2108-07732}, and MBPP-ET~\cite{DBLP:journals/corr/abs-2301-09043}). The human evaluation results show that {\tool} {elevates the performance (Pass@1) of GPT-4 on MBPP-sanitized from 70.96\% to 80.8\%, improves the performance (Pass@1) of ChatGPT on MBPP-ET from 51.52\% to 60.19\%.}
Besides, due to requiring the involvement of human participants, evaluating {\tool} could be very expensive and hard to reproduce. To perform automated evaluations of {\tool} across different LLMs and benchmarks without requiring user participation, we introduce a high-fidelity simulation method to simulate user feedback. We then conduct comprehensive experiments on four benchmarks (HumanEval~\cite{DBLP:journals/corr/abs-2107-03374}, HumanEval-ET~\cite{DBLP:journals/corr/abs-2301-09043}, MBPP-sanitized, and MBPP-ET) using two state-of-the-art LLMs (i.e., GPT-4 and ChatGPT). The results demonstrate that, in comparison with the default GPT-4, {\tool} achieves an average improvement of 11.52\% across four benchmarks; in comparison with the default ChatGPT, {\tool} achieves an average improvement of 15.07\% on four benchmarks.
Our main contributions are outlined as follows:   
\begin{itemize}
    \item \textbf{Framework}: We propose a novel framework, named {\tool}, which enables LLMs to detect ambiguous requirements and formulate targeted clarifying questions. {\tool} refines the ambiguous requirements based on the answers to clarifying questions and further generates code solutions.

    \item \textbf{User Simulation}: We introduce a user simulation method for producing high-fidelity simulated answers to the clarifying questions, which facilitates automated evaluations of {\tool} across different LLMs and benchmarks, eliminating the necessity for direct user participation.
    
    \item \textbf{Evaluation}: We conduct extensive experiments on four widely-used benchmarks to show that, {\tool} achieves substantial improvements across different models and benchmarks. A human evaluation further confirms the significant potential of applying {\tool} in real-world practice.
    
    \item \textbf{Data}: publicly accessible dataset and source code~\cite{website} to facilitate the replication of our study and its application in extensive contexts.
\end{itemize}

In the rest of this paper, Section \ref{sec:related} introduces the background and related work. Section \ref{sec:method} elaborates our proposed framework {\tool}. Section \ref{sec:exp} presents the experimental setup. Section \ref{sec:result} illustrates the results and analysis. Section \ref{sec:discussion} discusses the benefits and limitations of {\tool} and threats to validity. Finally, Section \ref{sec:conclusion} summarizes this work.
 

%% file: sec/rw.tex
\section{Background and Related Work}
\label{sec:related}

\subsection{LLM-based Code Generation}

Code generation is a hot research topic for software engineering and artificial intelligence communities. 
Recently, many LLMs have been proposed for code generation. One class of models is the encoder-decoder models, e.g.,  PLBART \cite{DBLP:conf/naacl/AhmadCRC21}, CodeT5 \cite{DBLP:conf/emnlp/0034WJH21}, and AlphaCode \cite{DBLP:journals/corr/abs-2203-07814}, which generally encode an input text into a context embedding and decode the embedding to a code solution. Another class of models is the decoder-only models that are trained with the next token prediction objective and generate code from left to right. GPT series models \cite{DBLP:journals/corr/abs-2107-03374, DBLP:journals/corr/abs-2204-06745}, PolyCoder \cite{DBLP:conf/pldi/Xu0NH22}, and InCoder \cite{DBLP:journals/corr/abs-2204-05999} are examples of such models. Among them, ChatGPT \cite{chatgpt} and GPT-4 \cite{DBLP:journals/corr/abs-2303-08774} are the state-of-the-art LLMs developed by OpenAI. They have demonstrated improved understanding and reasoning abilities, proficiency in comprehending the provided context, and the capacity to generate high-quality texts.

Since training or fine-tuning these LLMs is highly expensive, there has also been a lot of research focused on enhancing the performance of LLMs in code generation with minimal or no fine-tuning. Prompt Learning is one of the most important techniques for achieving this goal \cite{liu2023improving, DBLP:journals/corr/abs-2201-11903, DBLP:journals/corr/abs-2304-07590, nashid2023retrieval, DBLP:journals/corr/abs-2206-12839}. The Chain-of-Thought (CoT) \cite{DBLP:journals/corr/abs-2201-11903} is a novel prompting engineering technique, which can elicit LLMs to produce intermediate reasoning steps that lead to the final answer. It has shown impressive performance in complex reasoning tasks (e.g., arithmetic and symbolic reasoning) \cite{DBLP:journals/corr/abs-2201-11903, DBLP:journals/corr/abs-2205-11916}, and has therefore been applied to code generation \cite{li2023enabling, jiang2023self}. Inspired by CoT, Li et al. \cite{li2023enabling} propose a new prompting method, named Structured CoT (SCoT).  Different from CoT, SCoT explicitly introduces code structures and teaches LLMs to generate intermediate reasoning steps with program structures. Jiang et al. \cite{jiang2023self} propose a self-planning approach that can guide LLMs to understand code planning with few-shot demonstrations and write corresponding code planning for the given requirement. The aforementioned studies focus on leveraging and augmenting the reasoning capabilities of LLMs, that is, prompting LLMs to generate intermediate reasoning steps to enhance code generation performance. Nevertheless, they remain insufficient in addressing the ambiguous requirements provided by humans, as unclear user intent may mislead LLMs into producing incorrect reasoning steps, thereby yielding inaccurate results. Our {\tool} recognizes the importance of clarifying ambiguous requirements and proposes a novel framework that enables LLMs to automatically detect ambiguous requirements and ask targeted clarifying questions. By clarifying user requirements, {\tool} can generate code solutions that fulfill the user's intentions. GPT-Engineer \cite{GPT-Engineer} is a recent open-source Github repository. It utilizes manual-designed instructions to prompt LLMs to ask clarifying questions for the input user requirements, and then generates code snippets based on user feedback. However, GPT-Engineer asks clarifying questions for both ambiguous and unambiguous requirements, which is detrimental to the user experience and may result in incorrect code solutions \footnote{https://github.com/AntonOsika/gpt-engineer/issues/708}. While {\tool} can detect ambiguous requirements by checking whether the test outputs of sampled code solutions are identical. Furthermore, {\tool} employs prompting techniques to direct LLMs to first analyze the factors contributing to requirement ambiguity and then formulate targeted questions.

\subsection{Clarifying Question Generation}
The task of generating clarifying questions for ambiguous queries or dialogues has received much attention in information retrieval and dialogue system fields \cite{keyvan2022approach, trienes2019identifying, DBLP:journals/corr/abs-2008-07559, DBLP:conf/naacl/RaoD19, min2020ambigqa, krasheninnikov2022assistance}. In terms of information retrieval, many studies have pointed out that clarifying questions can help resolve ambiguous queries and improve user experience. For example, Wang and Li \cite{DBLP:conf/cikm/WangL21} find that search queries are often short and the underlying user intents are often ambiguous. They propose an effective template-guided clarifying question generation model, which employs Transformer to select a question template from a list of template candidates and fill in the question slot from a slot vocabulary. Eberhart and McMillan \cite{DBLP:conf/wcre/EberhartM22} propose a novel method to ask clarifying questions for query refinement, which utilizes a task extraction algorithm to identify query aspects and follows a rule-based procedure to generate questions.
In terms of the dialogue system domain, both rule-based and learning-based approaches have been proposed. Dhole \cite{DBLP:journals/corr/abs-2008-07559} proposes a novel method of generating discriminative questions by leveraging a simple rule-based system, which aims at seeking clarification from the user, thereby reducing the roboticity of the conversation and making the interaction considerably natural. Rao et al.~\cite{DBLP:conf/naacl/RaoD19} describe a method for generating clarifying questions, which uses a seq2seq model to generate a question given a context and utilizes another seq2seq model to generate an answer given the context and the question. 

In code generation, dealing with ambiguous user requirements has received little attention so far. To the best of our knowledge, Li et al. \cite{DBLP:conf/acl/LiMMG23} is the only research paper that addresses ambiguous requirements resolution for code generation. This work aims to clarify the ambiguous requirements missing key operations, e.g., API calls. It first collects a dataset named Code ClarQA containing natural language requirements, code, clarifying questions, and answers. Then, it proposes a code generation pipeline that can select relevant clarifying questions and their answers from the dataset for a given requirement for generating a code solution. However, the scope of applicability for this work is limited. Firstly, it primarily focuses on clarifying operational-level ambiguities, leaving other forms of ambiguity, such as semantic ambiguities in natural language requirements, less effectively addressed. Furthermore, it heavily relies on the constructed dataset, retrieving relevant questions for ambiguous requirements. If the dataset lacks similar requirements, the method's performance may suffer. Differing from this work, {\tool} is not limited to a specific type of ambiguous requirement clarification. And {\tool} can generate precise and targeted questions for various requirements by leveraging the powerful understanding ability of LLMs.

%% file: sec/method.tex
\section{Approach}
\label{sec:method}



\input{fig/overview_tool}

In this section, we introduce {\tool}, a code generation framework for LLMs. Figure \ref{fig:approach} illustrates the overview of {\tool}, which consists of four main stages: (1) \textbf{Test Input Generation} (Section~\ref{sec:method_1}), aiming at generating high-quality test inputs for a given requirement by using prompting techniques and heuristic mutations; (2) \textbf{Code Consistency Check} (Section~\ref{sec:method_2}), for leveraging the generated test inputs to conduct a consistency evaluation, and then identifying the ambiguous requirements; (3) \textbf{Reasoning based question generation} (Section~\ref{sec:method_3}), focused on generating targeted clarifying questions for the identified ambiguous requirements by prompting LLMs to engage in intermediate reasoning; (4) \textbf{Enhanced Code Generation} (Section~\ref{sec:method_4}), which incorporates the clarifying questions and their feedback to refine the original requirement and generate the final code solution based on the refined prompt. Below, we provide details for each stage in {\tool}.
 

\subsection{Test Input Generation}
\label{sec:method_1}
In this step, {\tool} aims to produce high-quality test inputs to effectively distinguish between code solutions with different functionalities.
There are many studies have attempted to employ LLMs for unit test case generation \cite{lemieux2023codamosa, schafer2023adaptive, vikram2023can} and have demonstrated impressive performance. Following prior work \cite{DBLP:journals/corr/abs-2305-01210}, {\tool} leverages LLMs as the test input generator and generates test inputs by adopting a two-step approach (i.e., seed input initialization and type-aware mutation). Specifically, {\tool} begins by designing a prompt to instruct an LLM in creating a set of seed inputs. It then performs type-aware mutations to generate a large number of new inputs. Our insights are: (1) on the one hand, since LLMs possess powerful understanding and reasoning abilities, using them as test input generators can produce high-quality inputs that remain valid even under semantic constraints. Take the example shown in Figure \ref{fig:case_study}, the problem in HumanEval requires the input string must contain balanced parentheses. Traditional input generators often face challenges in ensuring compliance with such semantic constraints. (2) on the other hand, LLMs are unsuitable for large amounts of automated test generation due to undesired speed and cost of querying such large models \cite{DBLP:journals/corr/abs-2305-01210}. Thus, we utilize a heuristic mutation-based method to accelerate the generation of numerous test cases, ensuring both stability and reliability.

\subsubsection{Seed Input Initialization}
{\tool} starts with designing a prompt for seed input initialization. As shown in Figure \ref{fig:prompt} (a), the prompt consists of three parts: (1) an instruction, designed to elicit LLMs to generate complex, difficult, and corner-case test inputs; (2) few-shot manually-crafted demonstrations, including a user requirement and ground-truth test inputs, which can assist LLMs in better understanding the task described in the instruction; (3) a query, for which LLMs generate input tests for based on it.
Specifically, we first finalize the prompt with the instruction, demonstrations, and the given requirement. Then, {\tool} utilizes the prompt to query LLMs to generate seed inputs. Finally, we collect these generated seed inputs to initialize a seed pool that will be used for mutation.

\input{fig/table1}

\subsubsection{Type-Aware Input Mutation}

After initializing a seed pool, {\tool} employs a type-aware input mutation strategy\cite{DBLP:journals/corr/abs-2305-01210} to generate higher-quality test inputs. Specifically, our approach follows the standard mutation-based fuzzing workflow \cite{DBLP:conf/issre/ZhangMZK11, fuzzing_lop}: (1) At each iteration, an input is randomly selected from the seed pool. (2) For the selected input, we inspect its data types and perform a single mutation operation consistent with its type to create a new test case. The basic mutations used for different types of inputs are illustrated in Figure~\ref{fig:mutation}. For simple data types, such as \textit{int} and \textit{float}, one mutation operation simply increases or decreases its value by 1. For compound types, we mutate the elements based on their internal types. (3) After completing a round of mutations, we add the newly generated inputs to the seed pool and repeat the aforementioned process until we attain the desired number of generated inputs.

\subsection{Code Consistency Check}
\label{sec:method_2}
A clear user requirement should be easy to understand and leave no room for interpretation, 
while an ambiguous user requirement can lead to stakeholders interpreting it in different ways. {Inspired by this, we make an assumption that, for a given requirement, if an LLM generates numerous code solutions with different functionalities, it signifies that the requirement can lead to the LLM interpreting it in different ways. Consequently, such a requirement necessitates further clarification and refinement.
In light of this assumption, we propose a simple yet efficient method to determine ambiguous requirements. First, we feed a given requirement into an LLM to sample $n$ code solutions. 
Then, these code solutions are executed with test inputs generated in the previous step. We obtain the test outputs of these programs and {compare the test outputs to inspect whether they are identical}.  If the outputs are identical, {\tool} considers these code solutions as interpreting the requirement in the same way, thus identifying the requirement as unambiguous. In this case, one of the sampled codes would be output as the final code solution. However, if the outputs are not identical, {\tool} believes the LLM has different understandings of this requirement when it produces code solutions and identifies the requirement as ambiguous. For these ambiguous requirements, as shown in Figure \ref{fig:approach}, we perform the code clustering to divide these code solutions into several groups based on their test outputs. Subsequently, {\tool} randomly chooses one code solution from each group and feeds these inconsistent code solutions into the next component to synthesize the prompt used for asking questions.

\input{fig/prompts_Detail}

\subsection{Reasoning Based Question Generation}
\label{sec:method_3}
Targeted clarifying questions facilitate users in articulating their intentions with clarity, ensuring that the responses obtained are directly pertinent to the unclear parts within the requirements. Vague or broad questions increase the risk of getting off-topic or irrelevant responses, which may hurt the performance of code generation. Therefore, upon identifying ambiguous requirements, it becomes essential to empower LLMs with the capability to craft precise and targeted questions. To achieve this objective, we devise a reasoning-based prompt aimed at directing LLMs to initially scrutinize the factors contributing to the ambiguity of the requirement and subsequently formulate targeted questions grounded in the analysis. The designed prompt is depicted in Figure \ref{fig:prompt} (b). It includes three parts: (1) an instruction, which describes the task (i.e., clarifying question generation) we want the LLMs to solve; (2) few-shot <requirement, inconsistent solutions, clarifying questions> triples as demonstrations, which help LLMs in understanding and solving the task; (3) a query, containing a user requirement and its code solutions, which is fed to LLMs for generating questions.

Specifically, {\tool} constructs the prompt to direct LLMs to analyze the factors contributing to the unclear requirement by understanding the functionalities of these inconsistent code solutions and comparing their disparities. The motivation is that, in software development, code solutions represent the specific implementation of requirements. If a requirement is ambiguous, different developers may have different interpretations and consequently write different code. Some of these inconsistent code solutions are incorrect or not in line with the original intent. By comparing these different code implementations, potential points of ambiguity in the requirements can be easily identified. After detecting the points of ambiguity in the requirements, the LLMs continue to generate targeted clarification questions based on the detection results.
 
Our proposed prompting shares a similar idea with the Chain of Thought (CoT)~\cite{DBLP:journals/corr/abs-2201-11903} prompting, which elicits LLMs to generate intermediate reasoning steps (analysis of the factors contributing to ambiguity) first, and then produce final results (targeted clarifying questions) based on these intermediate reasoning steps. In this way, {\tool} encourages LLMs to perform ``far-ahead planning'' \cite{bubeck2023sparks}, enabling them to better leverage their reasoning and comprehension abilities to enhance the quality of the generated questions.

\subsection{Enhanced Code Generation}
\label{sec:method_4}
Once user responses are captured, {\tool} combines them with the generated questions to refine the original requirement into a clear one. In particular, as shown in the query in Figure \ref{fig:prompt} (d), we pair each question and its corresponding answer to create a clarification, which are then appended to the end of the docstring to form the refined requirement. By refining an ambiguous requirement in this way, we can preserve the structural integrity of the docstring in the original requirement while enhancing it with additional clarifying information. Subsequently, we use the refined requirement to construct a prompt to instruct LLMs in generating the final code solution. The constructed prompt also consists of three parts, i.e., an instruction, some demonstrations, and a query, as depicted in Figure \ref{fig:prompt} (d).




%% file: fig/overview_tool.tex
\begin{figure*}[t]
\centering
\includegraphics[width=\textwidth]{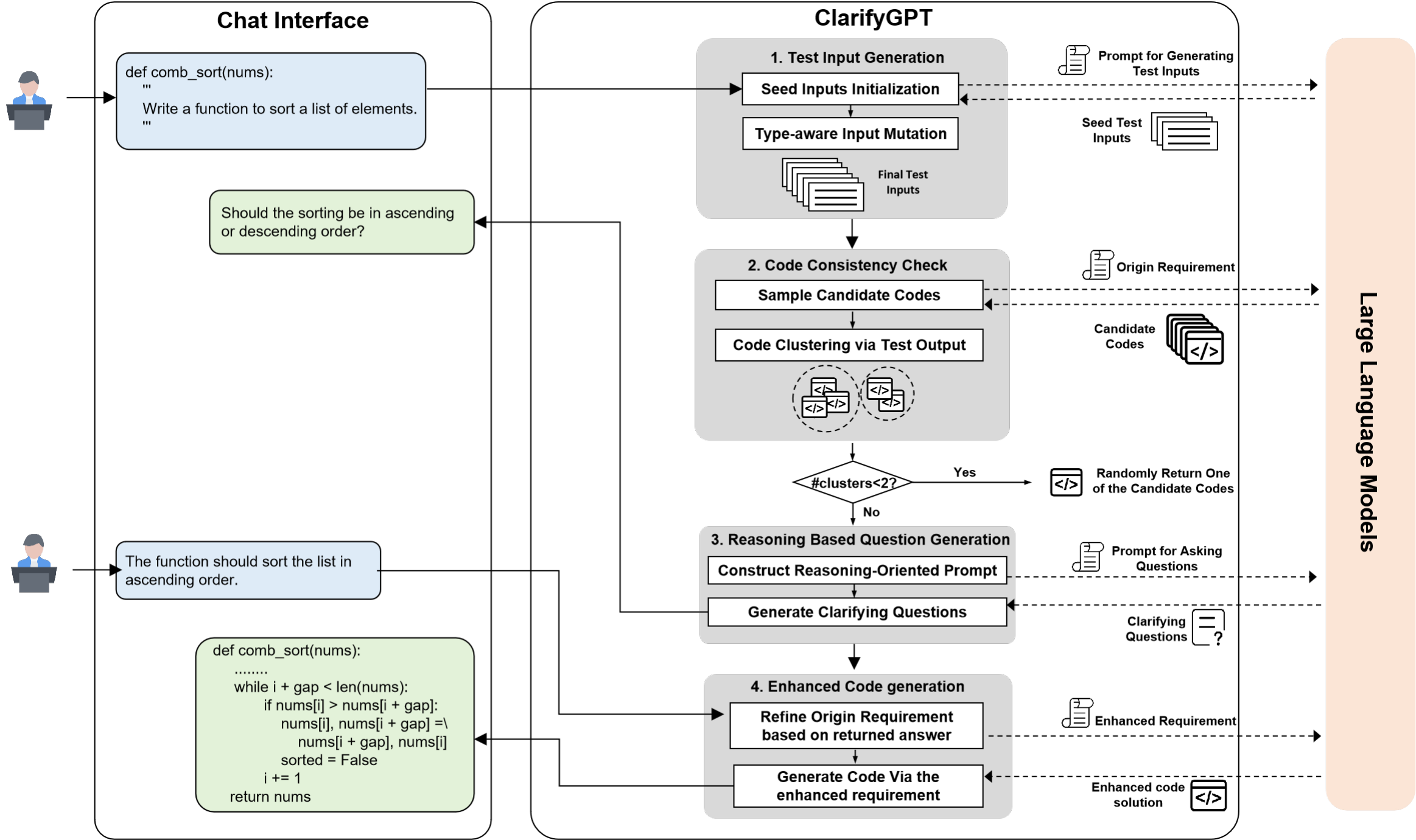}
\caption{The Overview of {\tool}}
\label{fig:approach}   
\end{figure*}

%% file: fig/table1.tex
\begin{figure*}[t]
\centering
\includegraphics[width=\textwidth]{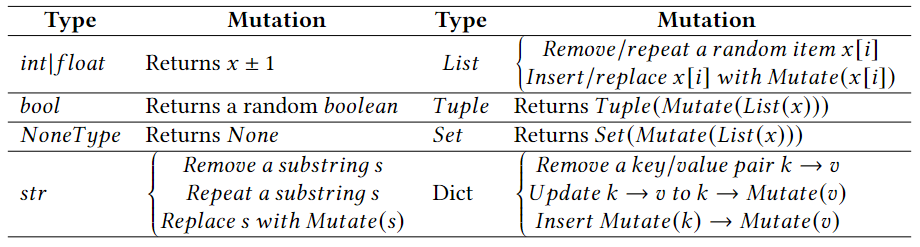}
\caption{List of basic type-aware mutations over input $x$ \cite{DBLP:journals/corr/abs-2305-01210}}
\label{fig:mutation}
\end{figure*}

%% file: fig/prompts_Detail.tex
\begin{figure*}[t]
\centering
\includegraphics[width=\textwidth]{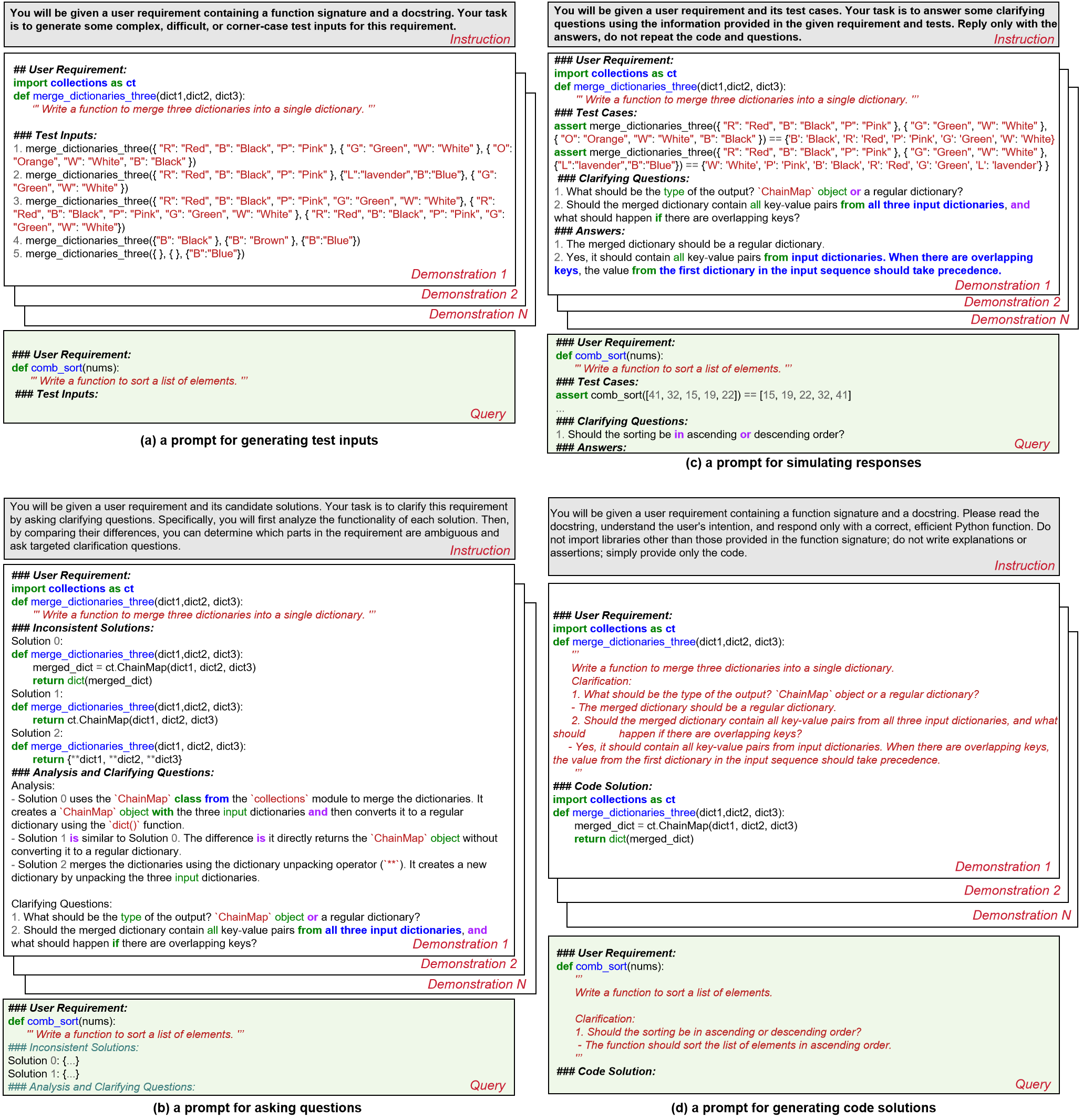}
\caption{The details of the prompts used in {\tool}} 
\label{fig:prompt}
\end{figure*}

%% file: sec/exp.tex
\section{Experimental Design}
\label{sec:exp} 
To evaluate the effectiveness of {\tool}, we conduct comprehensive experiments. In this section, we illustrate our experimental design, including research questions, models, benchmarks, metrics, baselines, and implementation details.

\subsection{Research Questions}
We address the following three research questions to assess the performance of {\tool}.

\textbf{RQ1: How does the {\tool} perform when receiving real user feedback in comparison to baseline approaches?}
In real-world scenarios, our {\tool} assists users in writing code by interacting with them, i.e., asking for clarification and receiving user feedback. Thus, in this RQ, we explore whether {\tool} with human in the loop can achieve higher performance than existing code generation baselines. Since evaluating interactive code generation with human participants is costly, we only select GPT-4 as the base model, and hire ten participants (including academic researchers and industry developers) to manually answer the clarifying questions generated by {\tool}. We compare {\tool} to three baselines on two benchmarks (i.e., MBPP-sanitized and MBPP-ET).

\textbf{RQ2: How does the {\tool} perform when receiving simulated user feedback compared to the state-of-the-art baseline approaches?} 
This RQ performs large-scale automated evaluations of {\tool} across different LLMs and benchmarks without requiring user participation, which aims to further verify whether {\tool} can achieve higher performance than existing code generation baselines. We first propose a user simulation method that leverages LLMs to simulate user feedback. Then, we apply three baselines and {\tool} to two representative LLMs (i.e., GPT-4 and ChatGPT), and evaluate their performance on four widely-used benchmarks (i.e., HumanEval, MBPP-sanitized, HumanEval-ET, and MBPP-ET).

\textbf{RQ3: How does the number of demonstrations in a prompt impact the performance of {\tool}?}
Prompting techniques could be sensitive to the number of demonstrations \cite{nie2022improving, gao2023constructing}. In this research question, we measure the performance of {\tool} with varying numbers of demonstrations to investigate the prompt robustness of {\tool}.

\subsection{Studied LLMs}
\label{exp:models}
There are many LLMs available for code generation. However, the specific context of this work necessitates that the LLMs possess a certain level of communicative competence, that is, the ability to comprehend human instructions and formulate clarifying questions. Thus, the LLMs without instruction tuning (e.g., InCoder~\cite{DBLP:journals/corr/abs-2204-05999} and CodeGen~\cite{nijkamp2022codegen}) are not suitable as the base models applied to {\tool} framework. In this work, we select two 
representative chat-LLMs (i.e., ChatGPT and GPT4) as base models to evaluate {\tool} framework.
\begin{itemize}[leftmargin=*]


\item{\textbf{ChatGPT}} 
{\cite{chatgpt} is one of the most powerful chat models empowered by OpenAI. It is trained using a novel approach called Reinforcement Learning from Human Feedback (RLHF), which seamlessly integrates reinforcement learning and human feedback. Specifically, ChatGPT is first trained with vast amounts of natural language text and code files. Then, it is fine-tuned through reinforcement learning, enabling it to adeptly comprehend and execute human instructions. In our experiments, We use OpenAI’s API to access the ChatGPT model, i.e., gpt-3.5-turbo.}

\item{\textbf{GPT-4}}~\cite{DBLP:journals/corr/abs-2303-08774} is OpenAI’s most advanced LLM, which can accept image and text inputs, emit text outputs. It is also trained with reinforcement learning and learns to follow human instructions. GPT-4 has demonstrated improved language understanding, allowing it to comprehend complex and nuanced contexts, making it highly effective on many downstream tasks, including text summarization, translation, and code generation \cite{bubeck2023sparks}. In our experiments, we use OpenAI’s API to access the GPT-4 model, i.e., gpt-4-turbo.

\end{itemize}

\subsection{Benchmarks}
\label{exp:benchmarks}

Following the previous work \cite{DBLP:journals/corr/abs-2303-06689, DBLP:journals/corr/abs-2207-10397, li2023enabling, DBLP:journals/corr/abs-2304-07590}, We conduct experiments on four public code generation benchmarks: HumanEval \cite{DBLP:journals/corr/abs-2107-03374}, MBPP-sanitized \cite{DBLP:journals/corr/abs-2108-07732}, along with their extended test case versions (i.e., HumanEval-ET and MBPP-ET \cite{DBLP:journals/corr/abs-2301-09043}). The statistics of these benchmarks are shown in Table~\ref{table:dataset}. 

\begin{itemize}[leftmargin=*]
\item{\textbf{HumanEval}~\cite{DBLP:journals/corr/abs-2107-03374}} is a hand-written problem-solving dataset crafted subsequent to the cut-off date of Codex's training dataset, consisting of 164 Python programming problems. Programming problems in the HumanEval concern language comprehension, algorithms, and mathematics. Each problem includes a function signature, a natural language requirement, and several unit tests. A problem is considered solved by code-LLMs when all unit tests are passed.

\item{\textbf{MBPP-sanitized}}~ \cite{DBLP:journals/corr/abs-2108-07732} is a hand-verified subset of MBPP (Mostly Basic Programming Problems) dataset, which contains 427 crowd-sourced Python programming problems, involving numeric manipulations, standard libraries functionality, and more. Each problem contains a function signature, a user requirement, and three test cases.

\item{\textbf{HumanEval-ET} and \textbf{MBPP-ET}}~\cite{DBLP:journals/corr/abs-2301-09043} are two extended versions of HumanEval and MBPP benchmarks with an average of 100+ additional test cases per problem. To improve the reliability of generated code evaluation, they collect many edge test cases that are not included in original benchmarks.

\end{itemize}

\input{tab/dataset_v2}

\subsection{Evaluation Metrics}
\label{exp:metric}
We evaluate the accuracy of the generated code using the metric Pass@$k$ \cite{DBLP:conf/nips/KulalPC0PAL19}. {This metric serves as an estimator of the generational capabilities under a specific budget, which is widely used in previous LLM-related studies \cite{DBLP:journals/corr/abs-2207-10397, DBLP:journals/corr/abs-2211-16490, DBLP:journals/corr/abs-2208-05950}}.
For each problem in the benchmarks, we generate $k$ code solutions, and if any of the $k$ code solutions passes all tests, this problem is considered solved. 
{In real-world development scenarios, generating $k$ code will impose a burden on developers, that is, they need to read and understand $k$ different code and select one as the target code. Thus, in this paper, the $k$ is set to 1, which satisfies most scenarios where developers consider only single-generated code~\cite{DBLP:journals/corr/abs-2304-07590, DBLP:journals/corr/abs-2303-06689}.} To avoid high variance and randomness, we run each approach three times and report the average results as the final results.

\subsection{Comparison Baselines}
\label{exp:baselines}

\begin{itemize}[leftmargin=*]
\item{\textbf{Default LLM:}} takes the original requirements directly from benchmarks as inputs to prompt LLMs for code generation.

\item{\textbf{CoT (Chain-of-Thought)}~\cite{DBLP:journals/corr/abs-2201-11903}}: generates a series of reasoning steps for each requirement by using the CoT prompt and then generates the corresponding code. To ensure the fairness of comparison, the CoT baseline has the same number of demonstrations (i.e., three demonstrations) and demonstration seeds.

\item{\textbf{GPT-Engineer}\footnote{https://github.com/AntonOsika/gpt-engineer}}: is a recent open-source Github repository. It utilizes manual-designed instructions to elicit LLMs to ask clarifying questions for the input user requirements and then generates code snippets based on user feedback. 

\end{itemize}

\subsection{Implementation Details}
The implementation details of constructing prompts and configuring models in {\tool} are as follows.

\textbf{Prompt Construction.} Since the four benchmarks do not have training sets, following previous work \cite{DBLP:journals/corr/abs-2201-11903, wang2022self}, we select the first three problems from each benchmark and extract the user requirements from these problems as demonstration seeds. Subsequently, we manually create distinct demonstrations for various prompts, as illustrated in Figure \ref{fig:prompt}. It should be noted that the reason we only create three demonstrations for each prompt is due to the input length limit of LLMs. 

\textbf{Model Configuration.} We treat the two LLMs used in the experiments as black box generators and only set a few interface parameters they provide without accessing internal parameters. For all LLMs, we set the \textit{top p} to 0.95, the \textit{frequency\_penalty} to 0. The \textit{max\_tokens} represents the maximum number of tokens to be generated, which is set to 800 for the prompt of asking clarifying questions and 300 for other prompts. In particular, we set the \textit{temperature} to 0, except when sampling code solutions, for which the \textit{temperature} is set to 0.8. We follow Chen et al. \cite{DBLP:journals/corr/abs-2107-03374} to truncate the content generated in HumanEval and MBPP by five stop sequences: ``\textbackslash nclass'', ``\textbackslash ndef'', ``\textbackslash n\#'', ``\textbackslash nif'', and ``\textbackslash nprint''.

%% file: tab/dataset_v2.tex
\begin{table*}[t!]
\renewcommand\arraystretch{1.3}
\centering
\caption{Statistics of benchmarks: the total number of problems in the benchmark (Problem Nums), the average number of test cases per problem (AVG.Tests per Problem), and the average/maximum/minimum number of prompt words in the benchmark (AVG/MAX/MIN.Words in Prompt).}
\label{table:dataset}
\resizebox{0.75\textwidth}{!}{
\begin{tabular}{c|c|c|c|c}
\hline
Benchmark             & HumanEval & HumanEval-ET  & MBPP-sanitized & MBPP-ET \\
\hline
Problem Nums          & 164    & 164       & 427   & 427      \\
\hline
AVG.Tests per Problem & 7.8    & 107.5       & 3.1   & 101.7      \\
AVG.Words in Prompt   & 67.7  & 67.7      & 14.5    & 14.5    \\
MAX.Words in Prompt   & 249  & 249       & 47     & 47          \\
MIN.Words in Prompt   & 17    & 17        & 7     & 7         \\
\hline
\end{tabular}
}
\end{table*}

%% file: sec/result.tex
\section{Results and Analysis}
\label{sec:result}

\subsection{RQ1: How does {\tool} perform when receiving real user feedback in comparison to baseline approaches?}

\noindent \textbf{Setup.}
In this RQ, we explore how {\tool} performs in real-world scenarios, that is, whether {\tool} can achieve higher performance than existing code generation baselines when receiving real user feedback. Specifically, we apply {\tool} to the GPT-4 model. Since MBPP-ET benchmark shares the same user requirements as MBPP-sanitized, we only apply {\tool} to the original versions of the benchmark (i.e., MBPP-sanitized) and report {\tool}'s performance on these two benchmarks using their respective unit tests. {\tool} first takes the user requirement of each problem in the benchmarks as input and determines them as ambiguous or unambiguous. Then, it generates clarifying questions for the ambiguous requirements (as shown in Figure \ref{fig:approach}). In total, we obtained 140 problems with ambiguous requirements from MBPP-sanitized benchmark. The average number of clarifying questions per problem is 2.85. We crafted three identical questionnaires for each problem, ensuring that each problem would be assessed by three different participants. Each questionnaire consists of three elements: (1) the (ambiguous) requirement of the problem, which describes the problem's intent; (2) the unit test cases containing expected input-output examples, which assist participants in understanding the problem's intent; (3) the generated clarifying questions, which participants are required to answer.

We recruited ten participants, including three Ph.D. students, two Master's students, two senior researchers, and three industry developers. None of them are co-authors of this paper. All participants have at least three years of experience in Python development, with six of them having more than five years of experience. Participants were initially provided with task descriptions and example questionnaires that contained appropriate question answers. After completing a training exercise, we assigned 42 problems to each participant and asked them to respond to the clarifying questions based on the information provided in the questionnaires. {Each problem will be solved by three participants.}

We collected the answers provided by the participants and input them into {\tool} to generate final code solutions. As mentioned earlier, we evaluated the correctness of the generated code on the two benchmarks using the unit test cases. Since each problem's clarifying questions were answered by three participants, we report the average Pass@1 results.

\input{tab/RQ2}
\noindent \textbf{Results.} The comparison results between the performance of {\tool} and other baselines are depicted in Table \ref{table:rq2}. The values in red are {\tool}'s relative improvements compared to the \textit{Default} baseline.

We can see that {\tool} (Human Feedback) achieves the highest performance on all four benchmarks. Compared with the \textit{Default}, {\tool} (Human Feedback) demonstrates superior performance with respect to the Pass@1 metric, achieving an increase of 13.87\% on MBPP-sanitized and 16.83\% on MBPP-ET. Furthermore, when compared to the best-performing baselines (i.e., CoT or GPT-Engineer), {\tool} (Human Feedback) also improves the performance of Pass@1 by 9.53\% on MBPP-sanitized and 9.52\% on MBPP-ET. This is mainly because {\tool} can proficiently identify ambiguous requirements and raise targeted clarification questions. Users easily clarify their intentions by responding to these questions, thus facilitating the generation of more correct code by LLMs. It indicates that {\tool}, as an interactive code generation framework, can support developers in writing code within real-world development contexts.


\mybox{
\textbf{Answering RQ1:} In human evaluation, {
{\tool} elevates the performance (Pass@1) of GPT-4 on MBPP-sanitized from 70.96\% to 80.8\%; elevates its performance on MBPP-ET from 51.52\% to 60.19\%. The relative improvement is 15.35\% on average, outperforming the baselines.}}

\subsection{RQ2: How does the {\tool} perform when receiving simulated user feedback compared to the state-of-the-art baseline approaches?}

\noindent \textbf{Setup.} Due to the involvement of human participants, evaluating the interactive code generation framework {\tool} is very expensive and hard to reproduce. A relatively simple solution is to conduct an offline evaluation \cite{aliannejadi2019asking}. However, it limits the system to selecting clarifying questions from a set of pre-defined or labeled questions, which does not transfer well to the practical development environment. In this RQ, we apply the User Simulation for Evaluation \cite{gordon1990evaluating, sekulic2022evaluating} method to facilitate automated evaluations of {\tool} across various LLMs and benchmarks, eliminating the necessity for direct user participation.

The most crucial aspect of simulating user feedback is to ensure that the created user feedback closely resembles the real feedback users would provide in the same environment. Low-fidelity simulations can result in {\tool} receiving feedback that is challenging to encounter in actual practice, thereby yielding misleading outcomes and impacting our evaluation of {\tool}'s performance. Hence, we propose a high-fidelity user simulation method that leverages LLMs to generate user responses by providing LLMs with clarifying questions and ground-truth test cases. Our key insight is that the ground-truth test cases contain expected input-output examples, reflecting the desired functionality sought by users. Endowing LLMs with this prior knowledge facilitates their understanding of user intent and enables the generation of high-fidelity simulated user feedback. To instruct LLMs to solve this task, we design a prompt (as shown in Figure \ref{fig:prompt}), which also consists of three parts: (1) an instruction, which describes the task (i.e., simulating the user responses) we want the LLMs to solve; (2) few-shot <requirement, ground-truth tests, clarifying questions, answers> quadruples as demonstrations, which help LLMs in understanding and solving the task; (3) a query, containing a user requirement and its ground-truth tests, which is feeds to LLMs for generating simulated responses.

We apply three baselines (Section \ref{exp:baselines}) and our {\tool} to two SOTA LLMs (Section \ref{exp:models}). We evaluate them on four benchmarks (Section \ref{exp:benchmarks}) and compare their performance by calculating the Pass@1 metric (Section \ref{exp:metric}). For a fair comparison, all baselines adopted the same experimental setup as our {\tool}.

\noindent \textbf{Results.}
\input{tab/main_result}
Table \ref{table:main_result} presents the comparison results between the performance of {\tool} (Simulated Feedback) and other baselines in terms of code generation. The values in red are {\tool} (Simulated Feedback)'s relative improvements compared to the \textit{Default} baseline.

Overall, {\tool} (Simulated Feedback) can substantially improve the performance of code generation, achieving gains across different LLMs and datasets. For GPT-4 model, compared with the \textit{Default} baseline, {\tool} (Simulated Feedback) demonstrates notable improvements in Pass@1 performance, achieving an increase of 11.34\% on the HumanEval dataset, 10.35\% on HumanEval-ET, 10.89\% on MBPP-sanitized, and 13.49\% on MBPP-ET. For ChatGPT model, when compared to the \textit{Default} baseline, {\tool} (Simulated Feedback) improves the performance of Pass@1 by 15.10\%, 13.12\%, 12.98\%, and 19.07\% on four benchmarks, respectively. The results demonstrate that {\tool}, which empowers LLMs to autonomously generate clarifying questions and refine user requirements based on user feedback, facilitates users in clarifying their intentions, thereby enhancing code generation performance by capturing user intentions.

We also note that, in comparison to the most related baseline (i.e., GPT-Engineer), exhibits superior performance with respect to the Pass@1 metric, achieving an average improvement of 11.45\%, 8.65\%, 6.95\%, and 8.56\% across the four benchmarks. We attribute the improvements to our novel techniques, i.e., ambiguous requirement identification and clarifying question generation. Posing clarifying questions for every user requirement results in needless LLM-Human interactions on unambiguous requirements, which places an additional burden on users and hurts the code generation performance when producing off-topic questions. While {\tool} can effectively identify ambiguous requirements without any supervised training by conducting the code consistency check. The inconsistent code snippets are taken as input to help {\tool} formulate targeted questions that guide users in clarifying ambiguity. 

Besides, we observe that the performance of {\tool} (Human Feedback) is slightly higher than that of {\tool} (Simulated Feedback). This suggests that our user simulation method may generate user responses that do not fulfill the users' intentions. However, both methods can significantly improve the performance of code generation and achieve consistent gains across different LLMs and benchmarks, demonstrating the reliability of our simulation method's evaluation results.

\mybox{
\textbf{Answering RQ2:} {\tool} (Simulated Feedback) improves the average performance (Pass@1) of GPT-4 across four benchmarks from 68.02\% to 75.75\%, improves the average performance of ChatGPT across four benchmarks from 58.55\% to 67.22\%. 
{Their relative improvements are 11.52\% and 15.07\%  respectively, and the average improvement is 13.27\%.}
}

\subsection{RQ3: How does the number of demonstrations in a prompt impact the performance of {\tool}?}

\noindent \textbf{Setup.}
In this RQ, we investigate whether the increase or decrease in the number of demonstrations will affect the performance of {\tool} (Simulated Feedback) on the code generation task. Specifically, due to the limitation of the input length of LLMs, we vary the number of demonstrations in the prompt from zero to three. Then, we apply the two LLMs to {\tool} and its variants, and assess their performance of four benchmarks. We run these methods three times and report the average Pass@1 results as the final reports.

\noindent \textbf{Results.}
Table \ref{table:RQ3} presents a comparison of the performance between {\tool} and its variants. Overall, {\tool} demonstrates robustness to the number of demonstrations in the prompts. When varying the number of demonstrations from zero to three, {\tool} consistently outperforms the Default baseline across two LLMs and four benchmarks.

We can observe that, as expected, the performance of {\tool} increases with the number of demonstrations. In particular, as the number of demonstrations in the prompt is incremented from zero to three, concerning ChatGPT, {\tool} achieves an average performance increase from 59.77\% to 67.22\% across four benchmarks. For the GPT-4 model, {\tool}'s average performance increases from 68.59\% to 75.75\%. This is mainly because more demonstrations can provide a variety of situations and information to LLMs, enabling them to better comprehend the context of the problem and the required solution. Furthermore, LLMs can learn to generalize better through demonstrations, that is, to infer a solution to a new situation from a known demonstration. This allows LLMs to better adapt to different inputs and requirements.
\input{tab/RQ3}

We also find that {\tool}'s performance in the zero-shot setting exhibits a marginal improvement over the Default baseline, while its performance in the one-shot setting is significantly enhanced compared to that of the Default baseline.   We attribute this difference to the fact that in the zero-shot setting, {\tool} is expected to generate meaningful responses without any demonstrations, which can be particularly challenging for complex tasks (e.g., requiring LLMs to generate targeted clarifying questions). What's more, zero-shot prompting relies solely on LLMs' pre-trained knowledge and the wording of the given prompts, which may not offer sufficient guidance or constraints for LLMs to produce accurate or contextually relevant responses. In contrast, the performance of {\tool} with the one-shot setting is significantly higher than that in the zero-shot setting and is close to the performance of {\tool} with the three-shot setting. This indicates that {\tool} has strong generalization performance when only one demonstration is provided. We believe that in practical usage scenarios, utilizing {\tool} in the one-shot setting can serve as a trade-off between effectiveness and efficiency.

\mybox{
\textbf{Answering RQ3:} Overall, {\tool} demonstrates robustness to the number of demonstrations in the prompts. When varying the number of demonstrations from zero to three, {\tool} consistently outperforms the Default baseline across two LLMs and four benchmarks.}


%% file: tab/RQ2.tex
\begin{table}
\centering
\renewcommand\arraystretch{1.3}
\caption{The Pass@1(\%) of {\tool} (Human Feedback) and baselines on two code generation benchmarks. {Numbers in \textcolor{red}{red} denote {\tool}'s relative improvements compared to the \textit{Default}.}}
\label{table:rq2}
\resizebox{0.7\textwidth}{!}{
\begin{tabular}{cccc} 
\toprule
\multirow{2}{*}{Methods}                                      & \multicolumn{3}{c}{GPT-4}                                                                       \\ 
\cmidrule(l){2-4}
                            & MBPP-sanitized        & MBPP-ET   & Average             \\ \hline
Default           & 70.96                 & 51.52     &   61.24          \\
CoT                & 72.68                 & 53.79     &   63.24          \\
GPT-Engineer      & 73.77                 & 54.96      &   64.37         \\ 
\hline
{\tool} (Human Feedback)              & 80.80                 & 60.19    &   70.50           \\
\textcolor{red}{Relative Improvement}  & \textcolor{red}{13.87\% ↑} & \textcolor{red}{16.83\% ↑}  & \textcolor{red}{15.35\% ↑} \\
\bottomrule
\end{tabular}
}
\end{table}

%% file: tab/main_result.tex
\begin{table*}[t!]
\renewcommand\arraystretch{1.3}
\centering
\caption{The Pass@1(\%) of {\tool} (Simulated Feedback) and baselines on four code generation benchmarks. {Numbers in \textcolor{red}{red} denote {\tool} (Simulated Feedback)'s relative improvements compared to the \textit{Default}.}}
\label{table:main_result}
\resizebox{\textwidth}{!}{
\begin{tabular}{ccccccc} 
\toprule
\multicolumn{2}{c}{Methods}                                              & HumanEval                                  & HumanEval-ET                               & MBPP-sanitized                             & MBPP-ET                                     & Average                                     \\ \hline
\multirow{5}{*}{ChatGPT} & Default                                       & 64.63                                      & 57.32                                      & 65.57                                      & 46.68                                       & 58.55                                       \\
                         & CoT                                           & 68.70                                      & 60.37                                      & 66.59                                      & 49.18                                       & 61.21                                       \\
                         & GPT-Engineer                                  & 66.26                                      & 59.76                                      & 69.09                                      & 50.20                                       & 61.33                                       \\ 
\cline{2-7}
                         & {\tool} (Simulated Feedback) & \textbf{74.39}            & \textbf{64.84}            & \textbf{74.08}            & \textbf{55.58}             & \textbf{67.22}             \\
                         & Relative Improvement                          & \textcolor{red}{15.10\% ↑} & \textcolor{red}{13.12\% ↑} & \textcolor{red}{12.98\% ↑} & \textcolor{red}{19.07\% ↑} & \textcolor{red}{15.07\% ↑}  \\ 
\hline
\multirow{6}{*}{GPT-4}   & Default                                       & 78.86                                      & 70.73                                      & 70.96                                      & 51.52                                       & 68.02                                       \\
                         & CoT                                           & 80.10                                      & 72.56                                      & 72.68                                      & 53.79                                       & 69.78                                       \\
                         & GPT-Engineer                                  & 79.27                                      & 71.75                                      & 73.77                                      & 54.96                                       & 69.94                                       \\ \cline{2-7}
                         & {\tool} (Human Feedback) & \textbackslash{}           &  \textbackslash{}        & \textbf{80.80}            & \textbf{60.19}             & \textbf{70.50}             \\
                         
                         & {\tool} (Simulated Feedback) & \textbf{87.80}            & \textbf{78.05}            & \textbf{78.69}            & \textbf{58.47}             & \textbf{75.75}             \\
                         & Relative Improvement                          & \textcolor{red}{11.34\% ↑} & \textcolor{red}{10.35\% ↑} & \textcolor{red}{10.89\% ↑} & \textcolor{red}{13.49\% ↑}  & \textcolor{red}{11.52\% ↑}  \\
\bottomrule
\end{tabular}
}
\end{table*}

%% file: tab/RQ3.tex
\begin{table*}[t!]
\renewcommand\arraystretch{1.3}
\centering
\caption{Experimental results of {\tool} with different number of demonstrations. {Numbers in \textcolor{red}{red} denote the relative improvement of {\tool} with different number of demonstrations compared to the \textit{Default}.}}
\label{table:RQ3}
\resizebox{\textwidth}{!}{
\begin{tabular}{ccccccc} 
\toprule
\multicolumn{2}{c}{Methods}                                      & HumanEval               & HumanEval-ET            & MBPP-sanitized          & MBPP-ET                 & Average                  \\ 
\hline
\multirow{5}{*}{ChatGPT} & Default                               & 64.63                   & 57.32                   & 65.57                   & 46.68                   & 58.55                    \\ \cline{2-7}
                         & {\tool} (zero-shot)  & 65.85  \textcolor{red}{1.9\% ↑}                  & 58.13  \textcolor{red}{1.4\% ↑}                 & 67.07  \textcolor{red}{2.3\% ↑}                 & 48.01  \textcolor{red}{2.8\% ↑}                 & 59.77 \textcolor{red}{2.1\% ↑}                   \\
                         & {\tool} (one-shot)   & 72.80  \textcolor{red}{12.6\% ↑}                 & 60.98 \textcolor{red}{6.4\% ↑}                  & 70.96 \textcolor{red}{8.2\% ↑}                  & 51.52 \textcolor{red}{10.4\% ↑}                  & 64.07  \textcolor{red}{9.4\% ↑}                  \\
                         & {\tool} (two-shot)   & 73.92 \textcolor{red}{14.4\% ↑}         & 63.21 \textcolor{red}{10.3\% ↑}                   & 72.60    \textcolor{red}{10.7\% ↑}                & 53.63  \textcolor{red}{14.9\% ↑}                  & 65.84       \textcolor{red}{12.6\% ↑}              \\
                         & {\tool} (three-shot) & \textbf{\textbf{74.39}} \textcolor{red}{15.1\% ↑} & \textbf{\textbf{64.84}} \textcolor{red}{13.1\% ↑} & \textbf{\textbf{74.08}} \textcolor{red}{13.0\% ↑} & \textbf{\textbf{55.58}} \textcolor{red}{19.1\% ↑} & \textbf{\textbf{67.22}}  \textcolor{red}{15.1\% ↑} \\ 
\hline
\multirow{5}{*}{GPT-4}   & Default                               & 78.86                   & 70.73                   & 70.96                   & 51.52                   & 68.02                    \\ \cline{2-7}
                         & {\tool} (zero-shot)  & 79.26  \textcolor{red}{0.5\% ↑}                 & 70.73 \textcolor{red}{0.0\% -}                  & 72.13 \textcolor{red}{1.6\% ↑}                  & 52.22 \textcolor{red}{1.4\% ↑}                  & 68.59  \textcolor{red}{0.8\% ↑}                  \\
                         & {\tool} (one-shot)   & 83.93 \textcolor{red}{6.4\% ↑}                  & 72.76  \textcolor{red}{2.9\% ↑}                 & 75.88   \textcolor{red}{6.9\% ↑}                & 55.97    \textcolor{red}{8.6\% ↑}               & 72.14  \textcolor{red}{6.2\% ↑}                  \\ 
                         & {\tool} (two-shot)   & 85.15  \textcolor{red}{8.0\% ↑}                    & 75.61    \textcolor{red}{6.9\% ↑}                  & 77.75   \textcolor{red}{9.6\% ↑}                   & 56.67    \textcolor{red}{10.0\% ↑}                  & 73.80 \textcolor{red}{8.6\% ↑}                      \\
                         & {\tool} (three-shot) & \textbf{\textbf{87.80}} \textcolor{red}{11.3\% ↑}   & \textbf{\textbf{78.05}} \textcolor{red}{10.3\% ↑}   & \textbf{\textbf{78.69}} \textcolor{red}{10.9\% ↑}   & \textbf{\textbf{58.47}}\textcolor{red}{13.5\% ↑}    & \textbf{\textbf{75.75}}  \textcolor{red}{11.5\% ↑}   \\
\bottomrule
\end{tabular}
}
\end{table*}

%% file: sec/discussion.tex
\section{Discussion}
\label{sec:discussion}
\subsection{Case Study}
\begin{figure}[htbp]
\centering
\includegraphics[width=\textwidth]{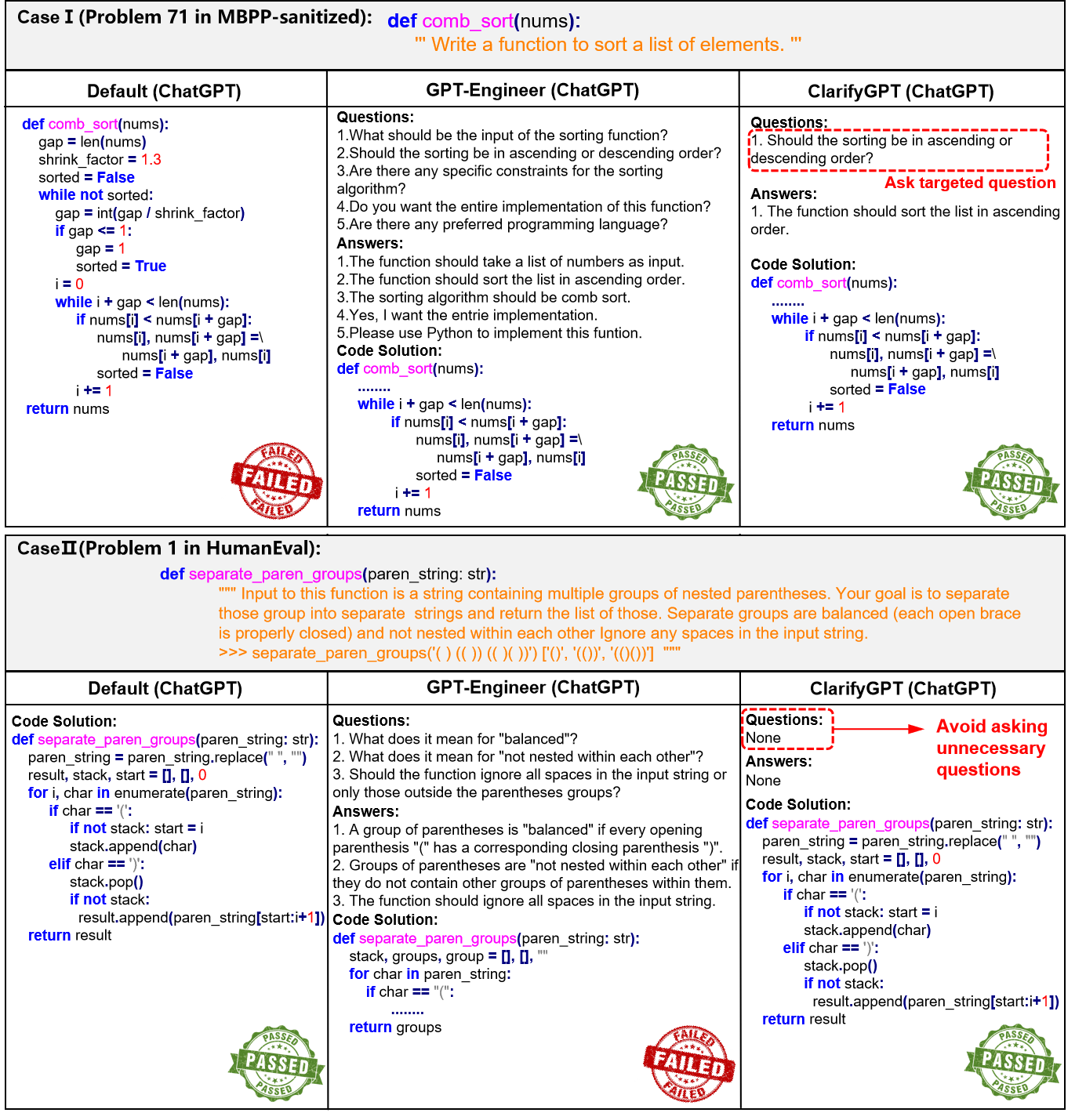}
\caption{{Two real cases from HumanEval and MBPP generated by two baselines and our {\tool}.}}
\label{fig:case_study}
\end{figure}
To further evaluate the effectiveness of our approach, we conduct a qualitative analysis. As shown in Fig. \ref{fig:case_study}, we select two representative examples from two popular code generation benchmarks (i.e., HumanEval and MBPP). Each input requirement consists of a function signature and an NL description. We take the ChatGPT~\cite{chatgpt} as the base model and utilize two baselines (i.e., Default and GPT-Engineer) and {\tool} to generate a code solution for each input requirement.

For the first example taken from MBPP-sa, the description \textit{``write a function to sort a list of elements''} does not specify whether this function should be sorted in ascending or descending order. The default ChatGPT directly generates a code solution that sorts the given list in descending order, which fails to pass the ground-truth test cases. GPT-Engineer poses five clarifying questions and generates a correct code solution based on the user responses. However, some of those questions are uninformative and can be answered by the information in the given requirement. For instance, the answer to the third question \textit{``Are there any specific constraints for the sorting algorithm?''} can be inferred by the function name \textit{comb\_sort} mentioned in the function signature. The fifth question \textit{``Are there any preferred programming language?''} also seems trivial, since we can easily know that the function should be implemented in Python based on the syntax of the function signature. Answering these questions cannot obtain additional information; instead, it generates superfluous dialogues that are detrimental to the user experience. Furthermore, it results in an increase in token counts for both the LLMs' input and output, consequently escalating operational expenses. By contrast, {\tool} can identify points of ambiguity in the requirements by comparing the different code implementations, thereby asking targeted clarifying questions. As a result, {\tool} only asks one question \textit{``Should the sorting be in ascending or descending order?''} and generates a correct code solution.

For the second example, the user requirement is well-defined. Default ChatGPT generates a correct solution, while GPT-Engineer produces an incorrect code solution. This discrepancy primarily arises from GPT-Engineer's inability to determine whether a requirement is ambiguous or not. Consequently, even for this unambiguous requirement, GPT-Engineer still poses three questions, which turn out to be uninformative. Moreover, these questions contribute to making the refined prompt excessively lengthy, potentially causing confusion for the LLMs. In contrast, our {\tool} can determine whether a requirement needs clarification by conducting a code consistency check. So we can see {\tool} does not raise any questions for this requirement but instead directly produces a correct solution.

\subsection{Benefits and Limitations}
In this section, we discuss some potential benefits and limitations of our {\tool}.

\textbf{Benefits.} (1) In contrast to prevailing LLM-based code generation methods \cite{DBLP:journals/corr/abs-2207-10397,DBLP:journals/corr/abs-2203-07814,DBLP:journals/corr/abs-2208-05950} that leverage post-processing techniques to sample a substantial pool of candidate codes and then select one, {\tool} aims to directly clarify the input requirements by asking clarifying questions. Hence, our framework contributes to the augmentation of interpretability in the code generated by LLMs. By clarifying specific details within the requirements or adding supplementary knowledge to them, users can readily discern corresponding alterations in the resulting code. This contributes to providing users with guidance on how to formulate requirements to improve code generation, thereby facilitating a clearer understanding of the generated code. (2) Our {\tool} improves the interactive skills of LLMs by empowering them with the ability to automatically ask clarifying questions for ambiguous requirements. In this way, it serves to facilitate users in identifying ambiguities within requirements and provides guidance in clarifying their intentions without requiring users to initially generate code and subsequently read and analyze code to refine requirements. Thus, {\tool} enhances the user experience and production efficiency.

\textbf{Limitations.} (1) Ideally, our framework is applicable to all LLMs. However, {\tool} necessitates that the LLMs possess a certain level of communicative competence, that is, the ability to comprehend human instructions and formulate clarifying questions. Thus, the LLMs applicable to our framework are limited, i.e., the LLMs without instruction tuning (e.g., InCoder~\cite{DBLP:journals/corr/abs-2204-05999} and CodeGen~\cite{nijkamp2022codegen}) are not suitable as the base models applied to {\tool} framework. (2) Due to the use of code consistency checking to determine whether a requirement needs clarification, {\tool} is required to generate test inputs for the requirement and compare the test outputs of the sampled solutions.  Therefore, {\tool} is not suitable for generating code with complex input (e.g., image or file). In addition, for some code that does not return output values (e.g., deep learning programs), using {\tool} may also be subject to some limitations.

\subsection{Threats to Validity}
\label{sec:discuss_threat}
The first threat to validity is the potential for data leakage. Since these LLMs are trained on open-source code repositories, it is possible that some public benchmarks were included in their training data. This could bias our assessment of the proposed approach, as some model outputs may be influenced by prior exposure to these benchmarks. To mitigate this threat, we carefully select HumanEval \cite{DBLP:journals/corr/abs-2107-03374}, MBPP-sanitized \cite{DBLP:journals/corr/abs-2108-07732}, and their respective extended versions for our evaluation. HumanEval is a manually crafted problem-solving dataset, introduced by OpenAI for assessing Codex's performance. MBPP-sanitized, on the other hand, is a hand-verified subset of the MBPP dataset, comprising 427 Python problems that have undergone crowd-sourced verification. These datasets have undergone meticulous manual review and have been widely employed in previous research studies \cite{DBLP:journals/corr/abs-2207-10397, DBLP:journals/corr/abs-2208-05950, DBLP:journals/corr/abs-2212-10481}.


The second threat to validity is the user simulation for evaluation. Due to the involvement of human participants, evaluating {\tool}, an interactive code generation framework, is very expensive and hard to reproduce. Thus, we propose a user simulation method to facilitate automated evaluations of {\tool} across various LLMs and benchmarks. However, low-fidelity simulations can result in {\tool} receiving feedback that is challenging to encounter in actual practice, thereby yielding misleading outcomes and impacting our evaluation of {\tool}'s performance. To mitigate this threat, we design a special prompt to provide LLMs with clarifying questions and ground-truth test cases. By endowing LLMs with this prior knowledge, {\tool} facilitates LLMs' understanding of user intent and enables the generation of high-fidelity simulated user feedback. The results show that the performance of {\tool} (Simulated Feedback) is very close to that of {\tool} (Human Feedback), proving that our proposed simulation method can serve as a good proxy for the automatic evaluation of {\tool}, eliminating the necessity for direct user participation.

The third threat pertains to the generalizability of our experimental results. To address this issue, on one hand, we have taken care to select two representative chat LLMs (ChatGPT and GPT-4) as our base models and four popular datasets as the evaluation subjects. We apply the two LLMs to our {\tool} and assess their performance on these four datasets. On the other hand, considering the inherent sensitivity of LLMs to prompts, we run baselines and {\tool} three times to help mitigate high variance and randomness. We report the average results as the final results. The results show that our {\tool} can substantially improve the performance of all LLMs, achieving consistent gains across different datasets. Therefore, we believe that {\tool} has good generalizability, and can perform effectively in many related contexts.


%% file: sec/conclusion.tex
\section{Conclusion}
\label{sec:conclusion}
In this paper, motivated by the observation that human developers typically ask clarifying questions when they are faced with ambiguous requirements, we argue that empowering LLMs with the ability to automatically clarify ambiguous requirements can improve code generation. To this end, we propose {\tool}, a code generation framework that enables LLMs to identify ambiguous requirements and generate targeted clarifying questions. 
Specifically, {\tool} consists of four main stages, i.e., test input generation, code consistency check, reasoning-based question generation, and enhanced code generation. For a given requirement, {\tool} first generates high-quality test inputs by using prompting techniques and heuristic mutations. Then, it utilizes the generated test inputs to conduct a consistency evaluation and identify the ambiguous requirements. Next, {\tool} formulates targeted clarifying questions for the identified ambiguous requirements by prompting LLMs to engage in intermediate reasoning. Finally, it incorporates the clarifying questions and their feedback to refine the original requirement and generate the final code solution based on the refined prompt.
In the evaluation part, we first apply GPT-4 to {\tool} and recruit ten participants to evaluate its performance on two public benchmarks. The human evaluation results show that {\tool} achieves a relative improvement of up to 16.83\% in Pass@1 compared to the \textit{Default} baseline. Additionally, to automate the evaluation of {\tool}, we introduce a high-fidelity simulation method to simulate user feedback. We conduct comprehensive experiments on four benchmarks (i.e., HumanEval, HumanEval-ET, MBPP-sanitized, and MBPP-ET) using two LLMs (i.e., GPT-4 and ChatGPT). The extensive results illustrate that {\tool} improves the average performance of GPT-4 across four benchmarks from 68.02\% to 75.75\%, and improves the average performance of ChatGPT across four benchmarks from 58.55\% to 67.22\%. Thus, we believe that {\tool} can significantly facilitate the practical application of LLMs in real-world development environments.